\documentclass{svproc}
\usepackage{url}
\usepackage{lscape}

\usepackage{longtable}
\usepackage{graphicx}
\usepackage{multirow}
\usepackage{cite}
\begin{document}
\mainmatter              
\title{Misinformation and its stakeholders in Europe: a web-based analysis}
\titlerunning{Misinformation in Europe}  
%
\author{Emmanouil Koulas\inst{1,2} \and Marios Anthopoulos\inst{2}
Sotiria Grammenou\inst{2} \and Christos Kaimakamis\inst{2} \and Konstantinos Kousaris\inst{2} \and Fotini-Rafailia Panavou\inst{2} \and
Orestis Piskioulis\inst{2} \and Syed Iftikhar Hussain Shah\inst{2} \and Vassilios Peristeras\inst{2,3}}

\authorrunning{Emmanouil Koulas et al.} 

\institute{University College London, London, United Kingdom,\\ Department of Computer Science\\
\email{emmanouil.koulas.20@ucl.ac.uk}
\and
International Hellenic University, Thessaloniki, Greece\\
School of Science And Technology\\
\email{\{ekoulas, manthopoulos, sgrammenou, ckaimakamis, kkousaris, fpanavou, opiskioulis, i.shah, v.peristeras\} @ihu.edu.gr}
\and
European Commission, Brussels, Belgium\\Directorate-General of Informatics}

\maketitle              

\begin{abstract}
The rise of the internet and computational power in recent years allowed for the exponential growth of misinformation phenomena. An issue that was a non-issue a decade ago, became a challenge for societal cohesion. The emergence of this new threat has led many stakeholders, especially in Europe, to act in order to tackle this phenomenon. This paper provides in its first part a literature review on misinformation in Europe, and in its second part a webometrics analysis on the identified key stakeholders. In the results we discuss who those stakeholders are, what actions do they perform to limit misinformation and whether those actions have an impact.
\end{abstract}
\section{Introduction}
Misinformation is the act of accidentally spreading false or inaccurate information \cite{online2019}. Some cases of Misinformation can be false rumors and pranks. Contrary to this, Disinformation describes the dissemination of malicious content, like, but not limited to, hoaxes, spear-phishing, and propaganda \cite{wooley}.

The problem of the quality of information and how it reaches the European citizens has become enormous nowadays. It is important to note that a false expressed opinion by anybody, even without an intention to manipulate facts, could potentially fuel disinformation. An even more significant fact, is that the structure of the internet allows for a snowballing effect, potentially reaching a huge audience \cite{renda}.

The phenomenon of Misinformation is emerging across Europe and has many manifestations. An insight into whether and to what extent European citizens trust their national information network comes from the fact that 23 out of the 28 EU states (82\%) have at least a medium level of trust in the information provided by their national media \cite{ebu}. Another statistic that should worry us regarding the confidence of European citizens in the media is that Social networks are the least trusted media in 29 out of 33 of the countries surveyed (88\%). Citizens of the EU are more likely to trust radio and television over the internet and social media \cite{ebu}. 

Not enough in-depth research has been conducted on networking patterns, which are a very valuable tool \cite{koulas2}. On a more specific level, few studies have been carried out about Misinformation and fake news spreading. Still, they primarily measured the level of the phenomenon and the corresponding impact of it on the political and social environment. In the domain our research was held, there aren't any relevant studies dealing with the correlation among the websites containing misinformation content. 

Nonetheless, few studies have focused on the analysis of Misinformation through Europe. A limited number of systematic reviews have examined the key actors that cause and encourage the spread of Misinformation. This study provides insights about the role of the European Governmental Organizations (GOs), Non-Governmental Organizations (NGOs), International Organizations (IOs), and International Institutes (IIs) in the spreading of that phenomenon. The method we used in order to construct our website collection sheet was by searching some key phrases about Misinformation on the search engines Google and Yahoo. On Bing, we faced some difficulties due to the lack of websites when we searched about specific key phrases. We collected and analyzed 49 seed websites with the criterion of the high ranking in the search engine. 

Our study provides a detailed webometric network analysis, based on the seed websites enlisted as Misinformation in Europe. It fills a gap in the literature for reviews of the correlation between sites. This study aims to empirically and methodologically assist in combating Misinformation both in Europe and the Global level. 

The remainder of the paper is structured as follows. In Section 2, we perform an extensive literature review, comprising of the misinformation and disinformation effects in Europe, related webometric studies and action taken by European stakeholder on the issue. In Section 3, we formulate our research questions and analyze the procedure we follow regarding the selection of the seed sites and their analysis. Results of the webometrics analysis follow at Section 4. Finally, in Section 5 we discuss our findings and we provide our concluding remarks in Section 6.

\section{Literature Review}

\subsection{Misinformation and Disinformation effects in Europe}
Misinformation and Disinformation are both phenomena that physically exist in everyday life almost since always. It is their digital nature that is novel nowadays and occurred a few years ago. 

Recently, the misinformation effect has been the source of significant concern in the Member States and has provoked discussions and investigations on this issue. The European Parliament, in June 2017, required, the European Commission analyzed thoroughly misinformation. Furthermore, it enquired from the Commission to formulating strategies for the effective mitigation of the problem. The Parliament even considered the possibility of legislative intervention to accomplish that mitigation, using as justification that fake news and Disinformation consist a considerable threat to the freedom of opinion, expression and democracy, which are of paramount importance under the European Union's Charter of Fundamental Rights. In a 2018 research conducted by the Robert Schuman Foundation, 68\% of European citizens claim they encounter fake news at least once a week, while at the same time, 37\% claim they encounter fake news daily \cite{schuman}.

The European Commission, in order to address the issue, on the 26th of April 2018, proceeded with the publication of a communication titled “Communication on tackling Disinformation: A European approach” \cite{commission}. Within this communication self-regulatory tools, which constitute the first step into countering effectively online disinformation in Europe, are contained.

President Juncker, in his speech on the State of Union, presented in detail the actions that will take place based on this Communication and noted that he would do everything within his power to protect the civil rights and democracy. The purpose of this communication is to ensure that all citizens can have access to objective, quality information.

The Communications work plan was a series of actions that lead to the creation of a robust mechanism that prevents fake news and wrong information to spread. A team of fact-checkers was created, who reviewed information coming from reputable public sources and evaluated them. Another task was to promote non-toxic and quality journalism and punish any news media channel that doesn't provide valid information to the public. 

At the same time, actions to educate people about choosing the right online and media information sources were taken in order to raise awareness for the issues of Misinformation and Disinformation and how these can affect the people. 

The most significant success of this Communication was the acceptance and adaptation of the Code of good practice to fight online disinformation \cite{commission1}, on the 26th of September 2018, which represents all the points and goals of the Communication. This Code successfully classifies all the prerequisites for a trustworthy only campaign, while at the same time enhancing fundamental principles, like the freedom of expression and media pluralism.

Misinformation can be tackled, and its effects can be mitigated through ICT and monitored via the use of the internet. Internet is a useful tool to accomplish that feat, and ICT is the medium that gives the capability to collect, accumulate, interpret, and show data in order to make crucial decisions and formulate strategies. And as the initiator of Europe's democratic system Robert Schuman has said: "Technology is changing, but our fundamental values remain. A citizen that is equipped with the necessary skills and that can listen, watch, and read critically is a prior condition for the success of these values".

\subsection{Webometrics related studies}
The application of bibliometric and infometric approaches to study the web, its information resources, structures, and technologies, is known as webometrics. The term webometrics is a coinage from the English word "web" and the andicet Greek "metric", which means to measure. Since the term was coined in 1997 by Tomas Almind and Peter Ingwersen, the value of webometrics quickly became established through the Web Impact Factor, the critical metric for measuring and analyzing website hyperlinks \cite{thelwall12}. If we would like to specify that methodology, an excellent definition to show would be the one by Mike Thelwall in 2004 "… the study of web-based content with primarily quantitative methods for social science research goals using techniques that are not specific to one field of study". The purpose of this alternative definition was not to replace the description within Information Science \cite{thelwall09}. The actual use of the first definition is to support the publishing of appropriate methods out of the scope of Information Science \cite{kunosic}.

Since the emergence of webometrics, this tool has become a useful methodology that applies in many fields such as the ranking of universities and scientometric evaluations or investigations of research areas. In order for the effective analysis of data for webometrics usage, it is of paramount importance to use known credible sources, for every category of webometrics.

The study from Roghaye Tafaroji, Iman Tahamtan, Masoud Roudbari, and Shahram Sedghi, which was conducted in 2012, aimed to present the findings of a webometric analysis of web sites of medical universities of Iran. They tried to investigate the Webometric ranking of Iranian Universities of Medical Sciences. In comparison to other similar studies that were conducted before and used inlinks and size as the main webometric criteria. The authors of this study examined rich text format files (doc, pdf, etc.) as a webometric indicator, and measured the impact of this new indicator on webometric ranks. The main findings indicate that Iranian Medical Universities performed poorly in regard to number of webpages, external links, and rich files. This observation is very useful because it can explain the anemic presence of these universities on the web  \cite{tafaroji}.

Another study presented a ranking of Alternative Search Engines (ASEs). Using indicators to evaluate a large amount of data that can be retrieved effortlessly and effectively, Bernd Markscheffel and Bastian Eine managed to create a picture of the most popular ASEs currently available. This approach allows further investigation for other studies, exploring the dynamics of the search engine market, while at the same time assessing the categories of ASEs \cite{markscheffel}.

A recent study conducted by The University of Burdwan measured and gave a clear idea about the information provided by the websites of the 25 High courts using this time just Google Search engine in contrast with the previous studies. This paper highlighted the various web impact factors, scores, and ranking of the websites of high courts in India and the final results that gave did open the door to further studies of other new areas of the webometric analysis \cite{mahji}.

Another study also examined information originating from the website of 71 universities in Bangladesh. The results given indicate that the universities of Bangladesh do not have greater web visibility and, it is clear that these universities need to focus on several issues to increase the visibility of their websites \cite{islam}. 

Furthermore, Webometrics Analysis was also used to measure Web Impact Factor (WIF) of 8 National Libraries' websites in South Asian Countries. WIF provides tools for quantitative research for several categories, like ranking, evaluation, categorization and comparison of websites, both on top-level and sub-level domains. The results visualized that the National Library of India leads with the highest Domain Authority and Page Authority, and it is followed by the National Library of Sri Lanka and National Library of Bhutan among the other National Libraries websites. Users visit the websites of the National Libraries for their information needs \cite{verma}. 

Last but not least, one of the significant studies regarding webometrics is "Open Data in Nepal: A Webometric Analysis", which measures the impact of Open Data in the Nepalese cyber domain \cite{acharya}. Acharya and Park's study serves as a guide for this study.

Taking in mind the related studies above it is clear that Webometrics is a tool used in many studies to examine the World Wide Web and give specific results about the construction and use of information resources. This is the reason why we decided to use Webometrics analysis in order to search the web and examine Misinformation in Europe.

\subsection{Anti-misinformation stakeholders inside and outside Europe}
As the volume of information flowing on the internet snowballs, the phenomenon of Misinformation is becoming more and more intense. For this reason, in recent years, many Governmental Organizations (GOs), Non-Governmental Organizations (NGOs) and International Organizations (IOs), inside and outside of the European border, have been mobilized to deal with Misinformation.

\subsubsection{European Level}
The anti-misinformation concern of the European Commission increased after some cases of intense manipulation of the public opinion on political issues. These phenomena occurred during the U.S and French presidential elections (2016-2017), as well as the Italian constitutional referendum (2016) \cite{service}. After a two-day conference in Brussels, the European Commission concluded that expert's help is vital in order to combat Misinformation \cite{commission17}. 

Finally, a High-Level Expert Group (HELG) on Artificial Intelligence was formed in January 2018 to reduce Misinformation, fake news, and Disinformation at any level within Europe. A report containing the best strategies and solutions about every disinformation issue depended on the same set of fundamental principles, which was the main deliverable of the HELG \cite{connect}.

\subsubsection{Member states of European Union}

Besides the European Commission, there are also many mechanisms within the states of Europe, that try to combat Misinformation.

In Germany and Croatia, strict laws about Misinformation and hate speech were applied. Websites that would not comply with the law within a specific period after the warning would pay a considerable fine \cite{bbc1, vintof}. 

In Italy, a portal where people could report to the authorities, fake news occurrences, was set, but unfortunately, it didn't work rationally because of the lack of knowledge about fake news \cite{commissario}. However, when a man was sent to prison for using a false identity in TripAdvisor reviews, the government's intentions were very clear. After that, the country's communications authority released a report on Misinformation \cite{agcom}.

Sweden, Denmark, Belgium, and the Netherlands launched campaigns on websites and social media in order to inform the citizens about Misinformation and fake news, on the initiative of their governments (2018-2019) \cite{funke}. 

In Spain, when Russia was blamed for spreading Misinformation concerning the Catalan referendum by national authorities, an agreement was signed between the two of them, to create together a cybersecurity team to prevent Misinformation \cite{funke}. Moreover, a task force of about 100 officials was activated during the 2019 elections, with the aim to combat fake political posts, especially on social media.

In France, a very innovative law was set on the press, which gives the power to the authorities to do whatever they must do with sites that illegally use fake news and Misinformation, enabling them to shut them down.  However, this measure was criticized especially from opposition senators and journalists because according to what they say, it is against the principle of proportional justice and press freedom \cite{ricci}.

In Greece, there are many NGOs nowadays, dealing with the refugees coming from regions where the situation is turbulent. Because of the significant number of migrants, this situation has become very sensitive. As a result, Misinformation and fake news are spread both from people and the media very fast. In order to help the migrant, many mechanisms collaborated to protect refugees' rights and fight Misinformation about this topic. They also take care of their housing and education.  These mechanisms are the "United Nations High Commissioner for Refugees (UNHCR)", together with the "Emergency Support to Integration and Accommodation (ESTIA)" program,  funded by the European Union Civil Protection and Humanitarian Aid, and some other local NGOs.

In the United Kingdom, a parliamentary report was published with the purpose to enforce citizens to avoid fake news and misinformation spreading, because the country suffers from a democracy crisis since the idea of Brexit came to the surface \cite{waterson}. Furthermore, the National Security Communications Unit was launched with the task to fight Disinformation from authority people and others, after Russia got involved in Brexit by spreading fake news \cite{bbc2}.

\subsubsection{NGOs}
At the same time, even though the mobilization of governmental organizations is essential, the contribution of non-governmental organizations to combating misinformation is just as remarkable. An international NGO, "Reporters Without Borders (RSF)", launched an innovative media to deal with Disinformation online \cite{boulay}. It is called the "Journalism Trust Initiative (JTI)" and is designed to encourage journalism by applying some agreed transparency standards to protect information and combat misinformation.

Moreover, charity NGOs, who are dealing either with refugees or with citizens that need help, are those who try besides their actions to protect people's rights and to publish only the real thing about the issues in which they are involved. "ActionAid" is one fascinating example of international NGO which has already offered very much in this sector.

\subsubsection{Anti-misinformation stakeholders outside Europe}

In addition to the efforts made to combat Misinformation within the European Union, efforts are also being made from countries outside of Europe. In some cases, Europe is firmly connected and immediately affected by the efforts made to fight Misinformation outside of Europe.

Russia has been one very important player in spreading fake news in recent years across Europe and the whole world, at both a political and social level. Especially in the US presidential elections, many campaigns where bots were used, were set in order to manipulate the result. For this reason, media continuously publishing fake news and misinformation are punished. For a start with a fine, if they would not conform with the law the people accountable for this would go to prison, and if that was not enough, their website would shut down \cite{stanglin}.

In the Americas, fines are the most widely used method of combating fake news and misinformation. In the USA, Brazil, and Chile, when someone is found to be disseminating fake news, the responsible party will be fined, and may even face prison time. This applies to everyone whether they are citizens writing on the net, whether they are journalists, webpages or even politicians who spread fake news \cite{funke}.

In Asia, many countries have adopted strict laws to deal with misinformation. In China and Malaysia spreading fake news is considered as a crime. Those who rumor fake news that can be harmful in public order are punished by the law with up to seven years in prison or public surveillance \cite{zhang, leong}. Moreover, south Asia countries, Thailand and Indonesia have also enforced laws in cases of misinformation detection. Many people were arrested for fabricating fake news, especially on social media, and many others paid huge fines \cite{funke}.

\section{Methods}
\subsection{Goals and Research Questions}
The rise of the internet and computational power has allowed for the disproportional growth of misinformation phenomena in the last years. In this study we want to discuss the measures taken by stakeholders in Europe to tackle those incidents and assess their effectiveness. For this, we formulated two research questions: (RQ1) Which are the key stakeholders and how do they fight the phenomenon of misinformation? (RQ2) Do the actions of the key stakeholders have a palpable impact? The first question aims to mapping the key stakeholders, as well as, assess their actions and cooperation. The second question tries to showcase the impact, if any, those actions have.

\subsection{Data Collection}
Our method for gathering information and seed sites regarding Misinformation was Google's search engine, as well as, Bing and Yahoo search engines. Trying to have a variety in our results and have the whole idea of Misinformation, we used different keyword combinations to get more accurate results. These are the search queries we used: 

\begin{itemize}
\item Misinformation in Europe
\item Role of NGOs in Europe to tackle Misinformation
\item Governmental Organizations tackling Misinformation in the Europe Area
\item Expanding Misinformation in society
\item The Consequences of Misinformation
\item Fighting Misinformation
\item Misinformation in Belgium and
\item Misinformation Tackling
\end{itemize}

The search results included most of European and Global NGOs, IOs, GOs, and IIs. The homepages of these organizations were visited and read individually to assess the importance of researching the phenomenon of Misinformation in Europe. We opted to include Belgium as a separate search query, due to the fact that the European Union's present there has led to the creation of a variety of think tanks, NGOs and corporations.

Since the exposure to a large scale of disinformation is proliferating, fighting misinformation is a top priority for the European Commission. Therefore we emphasize identifying the key factors that encourage the spread of this phenomenon throughout Europe. It is highly essential to understand the role of all the European Stakeholders and Institutions, such as the European Commission, Council of Europe, etc., that are responsible for measuring all the appropriate criteria that give us a more analytic point of view about the issue. 

Furthermore, it is significant to measure the impact and the consequences of Misinformation to find new and more efficient ways to counter misleading information in the European Area. Subsequently, we should emphasize on social-economic elements that will be able to lead us to determine a general framework for the protection of information throughout Europe, in the same way, the General Data Protection Regulation has been defined.

We choose to collect data from three Top Level Domain Categories and the specific eight Country Code Top Level Domains. In this context, to make it more understandable, we must observe the following hierarchy tree: 

\begin{figure}[h]
\caption{Hierarchy tree}
\centering
\includegraphics[width=0.5\textwidth]{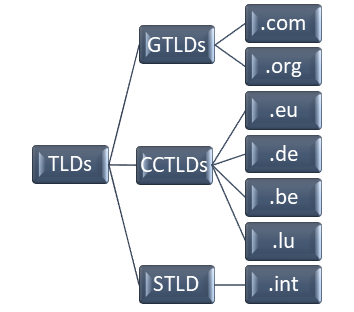}
\end{figure}

From the hierarchy tree in Figure 1, we notice the Top-Level Domain categories that there will be shown in the webometrics analysis and the following Country Code Top Level Domains. It's time to perceive the provenance of each domain. Firstly, .com derives from commercial, indicating its original intended purpose for domains registered by commercial organizations. Secondly, .org is truncated from organizations. Following the same way of thinking, we recognize that all Country Code Top Level Domains reserved for specific countries. In our case, we ended up with seedsites having the country codes of the European Union, Germany, Belgium, and Luxembourg. Finally, it's meaningful to recognize the role of Sponsored Top Level Domains. This category of the domain name is supported by a community or organization and considered to have the strictest application policies of all TLDs, as it implies that the holder is a subject of international law.

Table 2 lists the organizations, their established date, the sector in which they belong, their website address, and their URL.

\subsection{Process and Assessment}
The above websites are analyzed using Webometric Analyst 2.0 (http://lexiurl. wlv.ac.uk) and the Bing Application Program Interface, which is capable of carrying out advanced Boolean searches and tracking external websites linked to URL citations under study. Thus, the lists of external sites corresponding to the base query, i.e., the websites mentioned above, were obtained. 

Interlinkage and co-mention, as explained in Table 1, will be used for the data analysis.

\begin{table}[]
\caption{Analytical techniques and concepts of Webometrics\cite{acharya}}
\label{tab:tools}
\begin{tabular}{p{0.2\textwidth}p{0.8\textwidth}}
\hline
Inter-mention network analysis & Network diagrams illustrate the accompanying networks of the communication strength of the provided pairs of websites. It is the indicators based on asymmetric (directed) inter-mention counts between a pair of websites. A diagram illustrates the pattern of interconnectivity between collections of sites. This analysis gives a proxy for the hyperlinks between the websites under study \\
\\
Co-mention network analysis            & Network diagrams and their indicators based on the number of external sites referring to a pair of target sites. The co-mentions show something important in common but are not directly related to each other. The competitors who are also considered as stakeholders show a different pattern in the webometric analysis. Co-mention does not have a direction                                \\ \hline
\end{tabular}
\end{table}

For the assessment we are going to use the inlink degree, the outlink degree and the betweenness centrality. The inlink degree shows how many links from the network are inbound for each specific node, while the outdegree shows the outbound links towards other nodes of the network. Betweenness centrality shows the sum of times any particular node is found on the shortest path between different nodes of the network \cite{disney, valente, friedkin}. The higher the metrics, the more influential a node, thus an organization, is in the network. 
\section{Results}
 Figure 2 and Tables 2 and 3 answer our first research question. Figure 3 and Table 4 answer our second research question.  Figure 2 depicts a network diagram that demonstrates the inter-mention between websites conclusively. The red nodes and arrows show the linkage between the websites, whereas the green nodes indicate that there is no connectivity with any of the seed sites.
 
Every website domain and URL was converted accordingly to meet the requirements for Bing classification so that Bing can make the connections between the URLs and seed websites. According to the results below, websites are firmly connected and are the core of our findings, www.theverge.com, www.nytimes.com, www.cnet.com, www.washingtonpost.com, www.reuters.com, and www.bloomberg.com. As you see, the core URLs are oriented to the private company sector, and they have .com TLD. Most of the government based websites are connected, having a substantial presence on the web, for example, www.coe.int and www.poynter.org. 

\begin{figure}[h]
\caption{Inter-mention Network Diagram}
\centering
\includegraphics[width=0.95\textwidth]{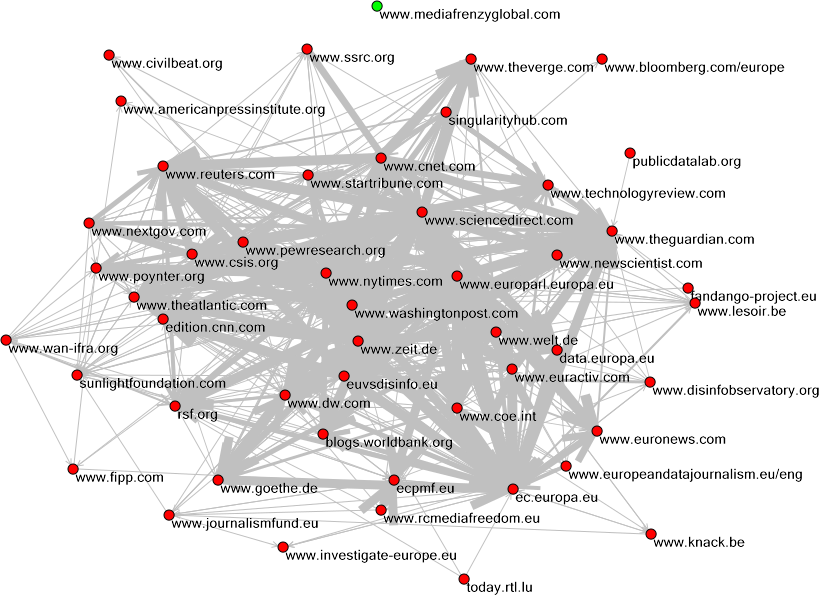}
\end{figure}

We also observe that www.goethe.de (which is a Nonprofit organization), www.zeit.de, and www.welt.de, which are both POs, are strongly connected to www.dw.com which is the most centralized website from all NPO's. Another seed site that has many connections to the central nodes of the diagram and vice versa is www.technologyreview.com and www.newscientist.com; both of them have a .com TLD. Additionally, we see that from websites with TLD .be and .lu only today.rtl.lu is not connected with any website. On the contrary, www.lesoir.be and www.knack.be are connected between them and also to the most central websites.

Furthermore, the interlinkage was investigated for the purpose of analyzing the online networking patterns in different networking scenarios. The values that we have observed are the networking density value for the directed network, which is 0,0791, and the value for the undirected network which is 0,1182, increased in comparison to the direct network. Density is calculated by diving the number of relations by the maximum number of possible relations. Density means an average value of entire cell blocks, when we refer to a network matrix that is weighted and valued. Next, 'degree' and 'betweenness' network centrality values are calculated. The term degree centrality refers to the amount of ties that are immediately connected to a node (i.e., website), rather than indirect ties to all others in the network\cite{acharya}. The two-degree centrality, specifically indegree, and outdegree, is calculated by the direction of the connection between two nodes. On the other hand, betweenness centrality measures how important a node is in the network. This is calculated by the effectiveness that a specific node plays as a broker, while connecting a pair of nodes. In this instance, the number of the shortest paths via the node is considered. Our network metrics were calculated utilizing built-in functions within Webometric Analyst.

\begingroup
\footnotesize
\begin{longtable}{p{0.1\textwidth}p{0.3\textwidth}p{0.1\textwidth}p{0.1\textwidth}p{0.3\textwidth}}
\caption{Description of the Selected Seed Sites}
\label{tab:seedsites}\\
\hline
No. & Organization Name                                                        & Est. date & Sector                                 & URL                                     \\ \hline
\endfirsthead
\multicolumn{5}{c}%
{{\bfseries Table \thetable\ continued from previous page}} \\
\hline
No. & Organization Name                                                        & Est. date & Sector                                 & URL                                     \\ \hline
\endhead
\hline
\endfoot
\endlastfoot
1   & Council of   Europe                                                      & 1999      & GOV                                    & coe.int                    \\
2   & European Commission                                                      & 1958      & GOV                                    & ec.europa.eu                   \\
3   & Public Data Lab                                                          & 2017      & IO                                     & publicdatalab.org               \\
4   & Reporters Without Borders                                                & 1985      & IO                                     & rsf.org                        \\
5   & European   External Action Service's East StratCom Task Force            & 2015      & NGO                                    & euvsdisinfo.eu                 \\
6   & Center for   Strategic and International Studies (CSIS)                  & 1962      & Non-profit Organization                 & csis.org                   \\
7   & Investigate Europe                                                       & 2014      & NGO                                    & investigate-europe.eu      \\
8   & Journalismfund.eu                                                        & 2008      & NGO                                    & journalismfund.eu          \\
9   & CNN Digital                                                              & 1980      & Private Company                        & edition.cnn.com                \\
10  & European Data   Journalism Network (EDJNet)                              & 2017      & Private Organization                   & europeandatajournalism.eu   \\
11  & European Centre   for Press and Media Freedom (ECPMF)                    & 2009      & NGO                                    & ecpmf.eu                       \\
12  & Poynter Institute                                                        & 1975      & GOV                                    & poynter.org                \\
13  & Social   Observatory for Disinformation and Social Media Analysis (SOMA) & 2018      & Project                                & disinfobservatory.org      \\
14  & Media Freedom Resource Centre                                            & 2015      & NGO                                    & rcmediafreedom.eu          \\
15  & WAN-IFRA - World   Association of News Publishers                        & 1948      & ORG                                    & wan-ifra.org               \\
16  & Parliament of Europe                                                     & 1952      & Int'l Institution              & europarl.europa.eu         \\
17  & EU Open Data Portal                                                      & 2012      & Portal                                 & data.europa.eu                  \\
18  & Euractiv                                                                 & 1999      & Network of Media                       & www.euractiv.com               \\
19  & Fandango Project                                                         & 2018      & Project                                & fandango-project.eu            \\
20  & The Guardian                                                             & 2011      & Private Company                        & www.theguardian.com            \\
21  & Sunlight Foundation                                                      & 2006      & Non-profit Organization                 & sunlightfoundation.com         \\
22  & New Scientist                                                            & 1956      & NGO                                    & newscientist.com           \\
23  & Fipp                                                                     & 1920      & Private Company                        & fipp.com                   \\
24  & The New York Times                                                       & 1851      & Private Company                        & nytimes.com                \\
25  & Washington Post                                                          & 1877      & Private Company                        & washingtonpost.com         \\
26  & Euronews                                                                 & 1993      & Portal                                 & euronews.com               \\
27  & MIT Technology Review                                                    & 1899      & Private Company                        & technologyreview.com       \\
28  & Media Frenzy Global                                                      & 2006      & Private Company                        & mediafrenzyglobal.com      \\
29  & Singularity University                                                   & 2008      & Univer-sity                             & singularityhub.com             \\
30  & The Atlantic                                                             & 1857      & Private Organization                   & theatlantic.com            \\
31  & The Verge                                                                & 2011      & NGO                                    & theverge.com               \\
32  & Civil Beat                                                               & 2010      & NGO & https://www.civilbeat.org/              \\
33  & The World Bank Group                                                     & 1944      & NGO                                    & blogs.worldbank.org            \\
34  & The Social   Science Research Council (SSRC)                             & 1923      & NGO                                    & ssrc.org                   \\
35  & Bloomberg                                                                & 1981      & Private Company                        & bloomberg.com/europe        \\
36  & Star Tribune                                                             & 1897      & Private Organization                   & startribune.com             \\
37  & Science Direct                                                           & 1997      & Private Organization                   & sciencedirect.com          \\
38  & NetGov                                                                   & 2008      & GOV                                    & nextgov.com                \\
39  & Lesoir                                                                   & 1928      & Private Organization                   & lesoir.be                   \\
40  & Knack                                                                    & 1971      & Private Organization                   & knack.be                    \\
41  & Deutsche Welle                                                           & 1953      & Non-profit Organization                 & dw.com                      \\
42  & Zeit                                                                     & 1946      & Private Organization                   & zeit.de                     \\
43  & Welt                                                                     & 1946      & Private Organization                   & welt.de                     \\
44  & Goethe Institute                                                         & 1951      & Non-profit                             & goethe.de                   \\
45  & RTL                                                                      & 1929      & Private Organization                   & today.rtl.lu                    \\
46  & American Press Institute                                                 & 1946      & Institute                              & americanpressinstitute.org \\
47  & Reuters                                                                  & 1851      & Private Organization                   & reuters.com                \\
48  & Computer Network                                                         & 1994      & Private Organization                   & cnet.com                   \\
49  & Pew Research Center                                                      & 2004      & Research Center                        & pewresearch.org            \\ \hline
\end{longtable}
\endgroup

Particularly, the 14 websites of our seed sites with the highest indegree and outdegree centrality are presented in Table 3. The Reuters (www.reuters.com) has the highest indegree centrality (74), and a Private Company 'The Guardian' (www.theguardian.com) has the highest outdegree centrality (66).  The balance between big organizations, like the European Parliament, and case specific seed sites like the EUvsDisinfo, ensure that the metrics are accurately depicting connectivity around misinformation.

\begin{table}[]
\caption{Top 14 websites with the highest indegree and outdegree centralities.}
\label{tab:centralities}
\resizebox{\textwidth}{!}{%
\begin{tabular}{lllllll}
\cline{1-7}
Organization            & Sector                  & Indegree &  & Organization            & Sector                  & Outdegree \\ \hline
Reuters                 & Private Company         & 74       &  & The Guardian            & Private Company         & 66        \\
Washington Post         & Private Company         & 54       &  & The New York Times      & Private Company         & 60        \\
The Atlantic            & Private Organisation    & 50       &  & Washington Post         & Private Company         & 56        \\
The Guardian            & Private Company         & 44       &  & The Atlantic            & Private Organization    & 28        \\
Bloomberg               & Private Company         & 36       &  & Reuters news agency     & Private Company         & 24        \\
The New York Times      & Private Company         & 26       &  & European Commission     & GOV                     & 24        \\
The Verge               & NGO                     & 24       &  & Science Direct          & Private Organisation    & 24        \\
Pew Research Center     & Research Center         & 20       &  & The Verge               & NGO                     & 20        \\
Deutsche Welle          & Non-profit Organisation & 16       &  & CNET (Computer Network) & Private Company         & 20        \\
CNET (Computer Network) & Private Company         & 14       &  & Deutsche Welle          & Non-profit Organisation & 14        \\
MIT Technology Review   & Private Company         & 14       &  & Poynter                 & GOV                     & 14        \\
New Scientist           & NGO                     & 14       &  & CNN World News          & Private Company         & 14        \\
The Star Tribune        & Online Media Company    & 10       &  & MIT Technology Review   & Private Company         & 12        \\
Euronews                & Portal                  & 10       &  & The Star Tribune        & Online media Company    & 10        \\ \hline
\end{tabular}%
}
\end{table}

Concerning the Private Organizations, like theguardian.com, theatlantic.com, washingtonpost.com, and reuters.com, they have high betweenness centrality (248, 185, 184, 180, 137). The local NGO www.theverge.com has the highest betweenness centrality among all NGOs (11,217). The Government Organization ec.europa.eu has a high betweenness centrality (113,867). The Nonprofit Organization Deutsche Welle (www.dw.com) have 42,35 betweenness centrality.  The betweenness centrality of a website shows the amount of control that this website exerts over the interactions of other websites in the network. The Pew Research Center is a research center that has the minimum betweenness centrality (0,2). Thus, it is noticeable that the Private Organizations have the highest betweenness centrality from NGOs.
In addition to the above conclusions, we see many websites with a weak connection between them such as www.disinfobservatory.org which is a technology-based company, also www.publicdatalab.org the only IO which has no connection. We have 6 NGO's websites with a minor presence on the web and no connection at all with the rest of the websites.  Finally, www.fipp.com and www.mediafrenzyglobal.com are two private companies with no connection between them.

Figure 3 shows the co-mention links of the websites. All of the nodes are colored red because all websites have at least one co-mention with another website. Each line's width is proportionately calculated and drawn based on the number of co-linking websites.

The paramount importance of the issues related to Misinformation is highlighted by the vast co-mentions among the websites analyzed. It is observed that the European Commission spearheads the efforts for tackling Misinformation.

The European Commission's role in promoting effectively European initiatives for tackling Misinformation can be shown by the fact that ec.europa.eu and www.disinfobservatory.org are co-mentioned 15 times, and ec.europa.eu and euvsdisinfo.eu are co-mentioned 185 times.

\begin{figure}[h]
\caption{Co-mention Network Diagram}
\centering
\includegraphics[width=0.95\textwidth]{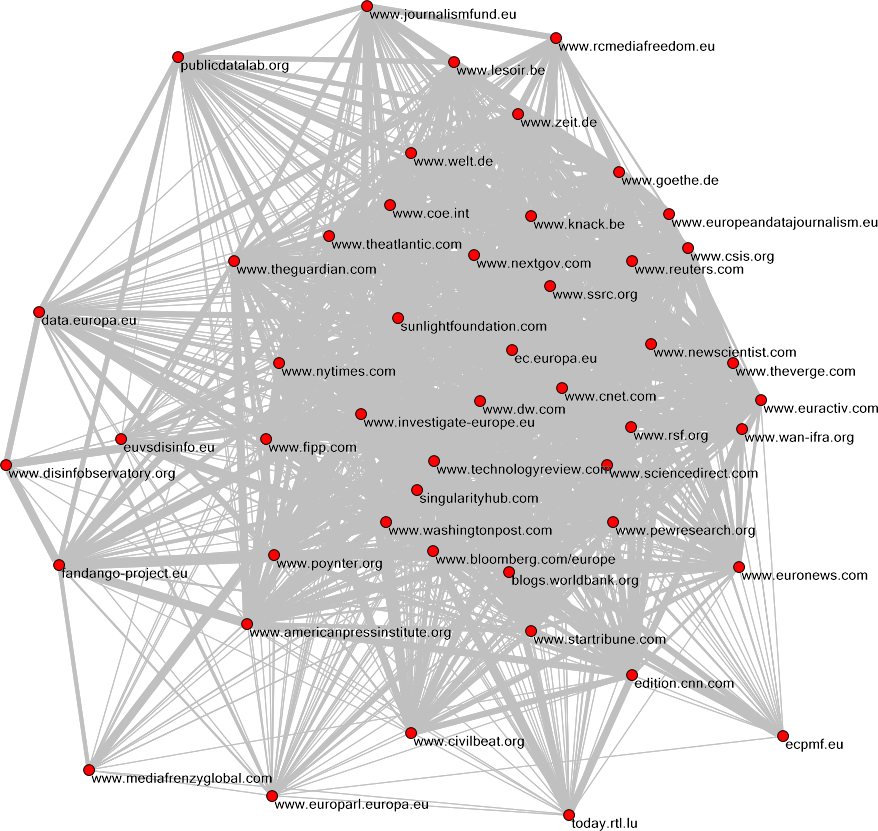}
\end{figure}

\begin{table}[]
\caption{Seed site calculation (N=49). Data values are defined as the total number of counts of TLDs citing seed sites.}
\label{tab:domains}
\begin{tabular}{llll}
\hline
No & Seed site TLDs/GTLDs/CCTLDs & Total Number of Seed Sites & Percentage (\%) \\ \hline
1  & .com                        & 21                         & 42.85\%         \\
2  & .org                        & 11                         & 22.44\%         \\
3  & .eu                         & 10                         & 20.40\%         \\
4  & .de                         & 3                          & 6.12\%          \\
5  & .be                         & 2                          & 4.08\%          \\
6  & .int                        & 1                          & 2.04\%          \\
7  & .lu                         & 1                          & 2.04\%          \\ \hline
\end{tabular}
\end{table}

At the same time, the Commission is used as a reference alongside the world's most influential think tanks, as it is observed that ec.europa.eu and www.theatlantic.com are co-mentioned 305 times; ec.europa.eu and www.csis.org are co-mentioned 359 times; ec.europa.eu and pewresearch.org are co-mentioned 541 times.

Lastly, the Commission is used to set a paradigm with other major organizations since ec.europa.eu and www.wan-ifra.org are co-mentioned 180 times; ec.europa.eu and www.rsf.org are co-mentioned 479 times, and ec.europa.eu and www.coe.int are co-mentioned 721 times.

Furthermore, it is observed that resource produced by European initiatives is often used alongside resources produced by leading think tanks. For example, euvsdisinfo.eu and www.csis.org are co-mentioned 49 times; euvsdisinfo.eu and www.pewresearch.org are co-mentioned 81 times, and euvsdisinfo.eu and www.theatlantic.com are co-mentioned 178 times. 

There are, however, initiatives that fall behind when it comes to European Projects. For instance, www.disinfobservatory.org and www.theatlantic.com are co-mentioned once, while www.disinfobservatory.org and www.csis.org are co-mentioned once, and www.disinfobservatory.org and pewresearch.org are co-mentioned twice.

Last but not least, European NGOs play an important role in the efforts to tackle misinformation since www.investigate-europe.eu and ec.europa.eu are co-mentioned 33 times; www.investigate-europe.eu and www.theatlantic.com are co-mentioned 661 times; www.investigate-europe.eu and www.csis.org are co-mentioned 841 times; www.investigate-europe.eu and www.pewresearch.org are co-mentioned 845 times; www.investigate-europe.eu and www.rsf.org are co-mentioned 849 times; and www.investigate-europe.eu and fandango-project.eu are co-mentioned 979 times.

All those findings show the massive effort, and the resources pooled from various stakeholders to tackle Misinformation.

\section{Discussion}

Misinformation is a constantly evolving threat that requires rigorous checks and balances in order to address it. The results can be categorized as those that stem from the theoretical part of the paper and those that derive from the webometrics analysis.

On the theoretical part, it is an interesting fact that in many cases, Webometrics is used as an evaluation system of a wide range of universities in the world. This system is known as a "ranking" system, where ranking describes a process where the position of the elements in a group regarding its entirety is defined by the relevance between the elements. The ranking process appears in many areas besides academics. For example, there was also a study on the ranking of Alternative Search Engines (ASEs) \cite{markscheffel}. 

Webometrics is a tool used in many studies to examine the World Wide Web for different reasons. In this project, we chose to use this tool to explore a totally different issue that also has an enormous impact on people and tries to interpret the data exported from the analysis. 

From the research on combating Misinformation at the state level, it can be concluded that the use of strict laws or regulations, to punish the people accountable for this phenomenon , is not an effective strategy. Inform people how to detect Misinformation, thus preventing them from reproducing it, is much more effective. Democratic societies ought to help their citizens learn how to acquire their information only from proven reputable sources, question what they read, examine its accuracy, avoid reading only the headlines of articles, and in case something is fake to avoid sharing it.

Last but not least, we observe the involvement of International Organizations in the effort to tackle Misinformation and Disinformation whose original mandate was totally irrelevant. For instance, the Council of Europe is heavily investing in the cyberspace domain, to remain relevant in a shifting world \cite{koulas}, while this change in perspective can foster meaningful collaborations \cite{brass}.

In the webometrics area, this study aims to investigate on a webometric dimension the role of all public and private sectors in Europe, as well as on the international level, that have taken actions as stakeholders to tackle the misinformation effect in Europe. We explore the structure of these stakeholders' portals and websites, their source, the organization's vision, methods of gathering and crosschecking data and information, and actions on the matter of Misinformation in Europe. We searched the seed sites countrywide and on an international level, for URL- based hyperlinks, title mentions and external links that refer to the seed sites of the stakeholders. According to the results, European governmental websites and portals are cooperating with the concerned NGOs inside Europe, but not as much as IOs are among themselves. In the co-mention network, all European portals show a strong connection with international websites and the other way around. Also, news organizations among them and other organizations of the same work- nature appear to have a very strong co-mention relationship. Besides, diverse organizations are also well co-mentioned. This could be because of the severe and sensitive nature of the issue and the urgency to counter it. Overall, European IOs and Governmental portals seem to have the most interlinkage and co-mentions as legal bodies that officially take actions to counter the Misinformation in Europe on a national governmental level with legislation, new laws and various efforts to raise awareness.

In our study we have identified 49 different stakeholder that took action in the fight against misinformation in Europe, as it is shown in Table 2. We found out that these efforts had some success, in terms of networking patterns. The limitations we faced are similar to those Acharya and Park \cite{acharya} face. Firstly, from a total of 71 possible websites, we manually selected the ones we deemed more important and more relevant to our study and used them as seed sites. Early tries to use all 71 of the original findings resulted in messy diagrams and due to the high number of irrelevant sites, the different metrics, especially the betweenness centrality, were boosted, without actually ensuring the misinformation focus we opted for. Secondly, many of our seed sites are not solely focused on Misinformation and Misinformation tackling alone but have made contributions towards that field. Thirdly, the seed site analysis was conducted with third party software and the search engine www.bing.com.

\section{Conclusions}
Misinformation is one of the significant challenges that modern societies need to address effectively because it severely impacts multiple aspects of our lives. We observe that there have been attempts to tackle Misinformation, with mixed results. The European Union spearheads these efforts, in cooperation with other organizations, but there it is possible to further enhance and improve those efforts. The ease world wide web provides for the spreading of Misinformation is a significant factor that increases the complexity of the situation. Further study should focus on the limitations of this study, as well as the usage of new technologies, like artificial intelligence, in order to process more data and yield better results.

%

\end{document}